\documentclass{article}

\usepackage[latin1]{inputenc}

\usepackage{hyperref} 
\usepackage{graphics} 
\usepackage{natbib}               
\usepackage{lineno}               

\usepackage{xspace}   
\usepackage{amsmath}  
\usepackage{amssymb}  
\usepackage{ifthen}   
\usepackage{graphics} 
\usepackage{tabularx} 
\usepackage{color}

\newcommand{\LUTN}[2]{\hbox{${#1}\mapsto{#2}\text{ \LUT}$}\xspace}
\newcommand{\LUT}{\textsc{lut}\xspace}
\newcommand{\LUTS}[1][6]{\ensuremath{{#1}\mapsto1\, \text{\LUT}}\xspace}

\newcommand{\DPA }{\textsc{dpa}\xspace}

\newcommand{\ACK}{\textit{Acknowledge}\xspace}
\newcommand{\ACKIN} {\ensuremath{{\mathcal S}_{in }}\xspace}
\newcommand{\ACKOUT}{\ensuremath{{\mathcal S}_{out}}\xspace}

\newcommand{\OOF}[1]{\textit{1-out-of-#1}\xspace}

\newcommand{\QPH} {\hbox{\textit{4-phase}}\xspace}
\newcommand{\DPH} {\hbox{\textit{2-phase}}\xspace}
\newcommand{\DPHE}{\hbox{\textit{2-phase-edge}}\xspace}
\newcommand{\DPHL}{\hbox{\textit{2-phase-ledr}}\xspace}
\newcommand{\QPHP} [1][]{\QPH\ protocol{#1}\xspace}
\newcommand{\DPHP} [1][]{\DPH\ protocol{#1}\xspace}
\newcommand{\DPHEP}{\DPHE\ protocol\xspace}
\newcommand{\DPHLP}{\DPHL\ protocol\xspace}

\newcommand{\PTY}[1]{\ensuremath{\phi({#1})}\xspace}

\newcommand{\LD}{\textbf{`2'}\xspace}                  
\newcommand{\LU}{\textbf{`1'}\xspace}                  
\newcommand{\LZ}{\textbf{`0'}\xspace}                  
\newcommand{\INV}{\ensuremath{\mathbf{\Omega}}\xspace} 

\newcommand{\VAL}{\textbf{ack}\xspace}
\newcommand{\Even}{\textbf{even}\xspace}
\newcommand{\Odd}{\textbf{odd}\xspace}

\newcommand{\VDD}{\ensuremath{V_{dd}}\xspace}
\newcommand{\VSS}{\ensuremath{V_{ss}}\xspace}

\newcommand{\RV}{\textit{rendez-vous}\xspace}
\newcommand{\CE}[1][]{C-element#1\xspace}

\newcommand{\OU}[2]{\ensuremath{\left[{#1}/{#2}\right]}\xspace}

\newcommand{\TBTDW}{\hbox{$2\times2-\text{\textsc{dw}}$}\xspace}                            
\newcommand{\TBODW}{\hbox{$2\times1-\text{\textsc{dw}}$}\xspace}

\newcommand{\gf}[1][]%
{\ensuremath{\ifthenelse{\equal{#1}{}}{\mathbb{F}_2}{\mathbb{F}_2^{#1}}}\xspace}

\newtheorem{remark}{Remark}

\newcommand{\REFFIG}[1]{Fig. \ref{#1}\xspace}
\newcommand{\REFEQ}[1]{Eq. \ref{#1}\xspace}

\newcommand{\EQIF}{\text{ if }}
\newcommand{\EQOT}[1][]{\text otherwise{#1}}

\begin{document}

\author{Ph. Hoogvorst \and S. Guilley \and S. Chaudhuri \and J.L. Danger \and T. Beyrouthy \and L. Fesquet}

\title{ A Reconfigurable Programmable Logic Block for a Multi-Style Asynchronous FPGA resistant to Side-Channel Attacks}

\date{Monday, October $31^{th}$, 2007}

\begin{abstract}

\emph{Side-channel attacks} are efficient attacks against cryptographic devices.         
They use only quantities observable from outside, such as the duration and the power consumption.

Attacks against synchronous devices using electric observations are
facilitated by the fact that all transitions occur simultaneously with some global clock signal.

Asynchronous control remove this synchronization and therefore makes it more
difficult for the attacker to insulate \emph{interesting intervals}. In addition the coding of
data in an asynchronous circuit is inherently more difficult to attack.

This article describes the Programmable Logic Block of an asynchronous
FPGA resistant against \emph{side-channel attacks}.
Additionally it can implement different styles of asynchronous control and
of data representation.

\end{abstract}


\section{Introduction} 
\label{introduction}

\textbf{\large Side-channel attacks} (SCA) have been put forward mainly by Paul Kocher \textit{et
al.} in 1996 in \cite{kocher-timing_attacks}. This first description
of a SCA explained how the mere observation of the duration of computations could allow an
attacker to retrieve the secret key. The attack was then improved and extended to other
cryptosystems~\cite{ches2000:timing-RSA-CRT,dhem98practical,hachez99timing,%
schindler-unleashing}.

In 1999 Kocher~\textit{et al.} described what they called ``\DPA\footnote{\textbf Differential
\textbf Power \textbf Analysis.}'' \cite{kocher99differential}. This new attack used the
power consumption instead to the duration but yielded the same result: the retrieval of
the secret key. The process of this latter attack is relatively simple: a large number of
cryptographic operations are monitored and the cipher text stored together with the
electric consumption. Then guesses were made of some parts of the secret key, which were
confirmed or or not by a statistical processing the data.
Other attacks against various cryptosystems were based on this method
\cite{ches2002:DPA-on-RSA-with-CRT,ches2002:DPA-of-OKEC,itoh-ches-2002}.

Countermeasures soon appeared to protect systems based on a strong algebraic
structure\cite{ches2002:ECC-with-CM,ches2002:DPA-counter-window,%
ches2001:algebraic-prot-DPA,ches2001:Jacobi-against-SPA-DPA}. At he
protection of opposite symmetric cryptosystems often consisted in introducing some
randomization either in the computing process or power consumption to prevent the
statistical processing of the acquired data. However ``counter-countermeasures'' also
appeared \cite{coron01boolean}. Some other protection schemes were designed
\cite{goubin-des,golic-multiplicative}.

An interesting and apparently efficient countermeasure is the WDDL\footnote{\textbf Wave
\textbf Dynamic \textbf Differential \textbf Logic.}~\cite{KABSP:2005} which duplicates each
signal in the circuit so that whatever the value is, one of the lines will toggle.
This countermeasure was enhanced by an improved routing of related signals
\cite{SHMP:2005}, which reduces the differences between the power consumptions of a '1'
and a '0'.

\medskip

\textbf{\large Asynchronous circuits}, the history of which dates back to 1950, are nowadays increasingly
considered as a viable alternative to classical synchronous designs. Indeed they feature
some very useful properties such as flexibility, robustness, high speed and low power.
This article brings another good reason to consider asynchronous designs: a greater
resistance against \textit{side-channel attacks}.

Some industrial applications of asynchronous ASIC and FPGA begin to appear both in the
academic world~\cite{MIPS1,Mar89,cornell} and in the industry~\cite{achronix}.

At the same time synchronous circuits are suffering from problems arising from the
distribution of the clock signal through the IC and the excessive power consumption (and
thus dissipation!).

As an asynchronous circuit has no centralized clock, the problems associated with the clock distribution,
clock skew and power consumption do not exist.
In addition this circuits offers advantages like:
\begin{itemize}
\item average-time performance,
\item lower electromagnetic radiation,
\item better robustness towards variations of the power voltage,
\item better robustness towards fabrication process variations~\cite{Nie94},
\item  better composability and modularity because of the simple handshake interfaces
       and the local timing~\cite{Spa01} and
\item better scrambling of the \textit{side-channel} information~\cite{anderson-async02,h2o-recosoc07,beyrouthy-icfpt07}. 
\end{itemize}
Asynchronous circuits thus seem to be a viable alternative which would remove these
limiting factors~\cite{Ren00}.

Due to these advantages, there has been a resurgence of interest in asynchronous design,
especially in the reprogrammable field.  There have been several recent successful design
projects such as ASPRO-216~\cite{Ren98}, AES crypto-processor~\cite{Bou05}, many of
Philips designs targeting low power~\cite{Ber94b,Kes97}, projects focused on designing an
asynchronous FPGA from a synchronous one, like MONTAGE~\cite{Hau92} and
PGA-STC~\cite{Mak98} or targeting asynchronous application-specific FPGAs, locally
synchronous, like GALSA~\cite{Gao96} and STACC~\cite{Pay97} or completely asynchronous
like PAPA~\cite{Tei03,Tei04}, and other recent works~\cite{Ho02,Fes05b,Fes06,Huo05,Fes05}.
PGA-STC was developed to implement two-phase bundled-data systems such as micro pipelines,
GALSA for massively parallel computing architecture, STACC for reconfigurable computation
and PAPA was mainly created and optimized for pipe-lined processes.

\textbf{This article} describe the design of the PLB%
\footnote{\textbf Programmable \textbf Logic \textbf Block.}
of a new asynchronous FPGA with security as the main requirement, even
at the expense of performance. Indeed in the particular case of cryptography performance
is second to security even if it cannot be ignored. The FPGA must be able to implement
various styles of asynchronous protocols and different representations of data so as to
enable comparisons between these representations and protocols as for their ability to
thwart the side-channel attacks.

Section \ref{asynchronous-representation} describes the representation of data and the
different asynchronous protocols used in the FPGA. We also discuss their suitablity for trusted computing. Section \ref{4-Phase-protocol} shows
the construction of the PLB to implement the 4-phase protocol using both binary and
ternary representations of data. Section \ref{2-Phase-protocol} shows the necessary additions to the PLB
to accommodate the 2-phase protocols. Section \ref{programming-the-fpga} shows how the FPGA is
programmed. Finally section \ref{conclusion} concludes the article.

\section{Asynchronous Representation of Signals}
\label{asynchronous-representation}

As opposed to synchronous data, whose validity is guaranteed by the timing of some global
``clock'' signal, the asynchronous computations are synchronized by the availability of data
and, when necessary, by a Request/Acknowledge handshake signalling.

A formal description of delay insensitive representation of data can be found
in~\cite{codes}. In the \textbf Quasi-\textbf{D}elay \textbf{I}nsensitive (QDI) protocols
the request is carried by the data itself. This allows to obtain a reliable design,
independent of the routing.

The data are transmitted together with the availability information and thus a logic signal or,
shorter a ``\textbf{signal}'', must be represented by more than a single electrical signal or,
shorter, a ``\textbf{wire}''\footnote{If one could work with non-standard electrical
levels, a $\{-5~V,0~V, +5~V\}$ representation on a single wire per signal would be
acceptable in some cases but we shall restrict ourselves in the following pages to
standard CMOS levels: \VDD and \VSS.}. In this article, a wire is able to take one of two
values, which we denote \textbf{0} and \textbf 1 regardless of their actual electric
implementation.

In order to avoid glitches, a sufficient condition is that \textit{given a signal $S$
represented by $n$ wires, the transmission of a new value of $S$ must consist in
\textbf{exactly one} of the $n$ wires changing its electrical state}. This means that the
number of wires is greater than or equal to the number of the states of $S$. As silicon
and routing is a precious resource, the number of wires representing a given signal will
thus be equal to the number of possible values of this signal.

The most frequently used kind of signal is the binary signal, which carries a \{\LU,\LZ\}
information. Such a signal is encoded with 2 wires. This representation is called
``\textbf{Dual-Rail}'' or ``\OOF 2''. However ternary signals, which carry a
\{\LZ,\LU,\LD\} information, can also be thought of. Such a signal is represented by 3
wires and one speaks of ``\OOF 3'' representation. This representation is more compact
than the \OOF{2} as for arithmetic: 6 wires in \OOF 2 represent 3 \OOF 2 signals which
can take 8 valid values, compared to two \OOF{3} signals, which can take 9 valid values.
However due to the greater complexity of gates in \OOF 3 representation, the binary
signals are most of the time preferred.

An asynchronous design may need additional signals, which are specialized to
synchronisation. These signal carry no \textit{data} information and can thus be coded on
a single wire. They will be referred to as \ACK signals. The inputs of the
gates which receive such a signal will be denoted \ACKIN and those driving these signals
will be called \ACKOUT.

\subsection{Asynchronous Protocols}
\label{PROTOCOLS}

There are two main families of QDI asynchronous communication protocols, which differ by
the nature of the signalling information: the \DPHP[s]  and the \QPHP[s].

\subsubsection{4-Phase Protocol}

Under a \QPHP, valid values of a signal are separated by a special value, denoted \INV.
The transmission of a value $x$ from an emitter to a receiver proceeds as follows:

\begin{center}
\begin{tabular}{|rlcl|}
	\hline
 &	Emitter      &                 & Receiver          \\
	\hline
1&	sends $x$    & $\longrightarrow$ &                   \\
2&	             & $\longleftarrow $ & acknowledges $x$  \\
3&	sends \INV   & $\longrightarrow$ &                   \\
4&	             & $\longleftarrow $ & acknowledges \INV \\
\hline
\end{tabular}
\end{center}

For instance, if a signal $S$ is represented by $n$ wires $(S_0, S_1,,...,S_{n-1})$, the
\INV value will be implemented as the $n$-tuple $(0,0,\ldots,0)$ while the value $i$ will
be represented by $(0,\ldots,,0,S_i=1,0,\ldots,0)$.

This particular kind of \QPHP is named ``\textsc{wchb}''\footnote{\textbf{W}eak
\textbf{C}ondition \textbf{H}alf \textbf{B}uffer.} in~\cite[Sec.~2.3.1]{phd_rigaud} and as
\textsc{dpl}\footnote{ \textbf{D}ual-\textbf{R}ail \textbf{P}recharge \textbf{L}ogic}
among the secure computing community~\cite{ches-PoppM05}.

\subsubsection{2-Phase Protocols}

Under a \DPHP, no special value is used to separate valid ones. The transmission of a
value $x$ from an emitter $E$ to a receiver $R$ proceeds as follows:

\begin{center}
\begin{tabular}{|rlcl|} \hline
 &	Emitter   &                 & Receiver            \\ \hline
1&	sends $x$ & $\longrightarrow$ &                   \\
2&	          & $\longleftarrow $ & acknowledges $x$  \\ \hline
\end{tabular}
\end{center}

In this article we will describe the implementations of two \DPHP[s]:
\begin{description}
	\item[\DPHEP:]\ \\
		a signal $S$, which can take $n$ values is represented by $n$ wires and the
		arrival of a new value $i$ is signalled by wire $i$
		toggling \hbox{$0\rightarrow1$} or \hbox{$1\rightarrow0$}.
		Note that the instantaneous values of the wires is not
		significant under this protocol: only the toggles are significant.
	\item[\DPHLP:]\ \\
		a signal $S$ is represented by two wires: $S_d$ and $S_r$. The arrival of a new
		value $x$, is signalled by one of $O_d$ and $O_r$ toggling \hbox{$0\rightarrow1$}
		or \hbox{$1\rightarrow0$} and the value is given by $O_d$. 

		Note that the requirement that any change of the value of the signal be
		implemented by the toggling of exactly 1 wire limits the \DPHLP to binary signals.
\end{description}

\begin{remark}
The \QPHP can be considered as a \DPHP in which all ``valid'' values are followed by a \INV
dummy value and in which the
gates return to the \INV value as soon as all inputs have received the \INV value. The
\DPHP[s] are thus inherently twice faster as the \QPH ones. This is especially important
in a FPGA, in which the routing delays are often the limiting factor of the speed of the
system. However, even if twice faster, they lead to much more complex gates than the
\QPH ones.
\end{remark}

\subsection{Initialization of the System}

At the initial time of the system's operation, all gates must be reset to a known,
deterministic value. (This is also true for synchronous systems even if some flip-flops
sometimes need no initialization.)

The requirement of \textit{a known, deterministic value}, implies no specific value to the wires.
However the  simplest initialization, which we shall use in this article, consists in
initializing all gates so that all wires be set to 0.

The consequence of this initialization is that the parity of the Hamming weight of any
signal is 0 just after reset, which implies that its parity is even.

The relevant property just after RESET is thus that:
\begin{itemize}
\item under a \QPHP an \INV value is thus output by all gates and
\item under a \DPHP the parity of the Hamming weight of the outputs of any gate is 0.
\end{itemize}

\subsection{Request Signalling}

The \textbf{Request} event is coded into the data of the QDI protocol itself;
a request corresponds to a change of one of the wires encoding the signal.
A gate will be ready to perform its computation when each of its input have received a
request and when all gates using its output have acknowledged the last value sent.

If performance were the major requirement this would not be true: for instance, a AND gate
could perfectly output a \LZ as soon as one of its inputs has received a \LZ.  But such an
early evaluation would occur only when some input(s) receive a \LZ and never when all
receive \LU. This difference in timing could potentially leak some information about the
computations being performed to a malevolent observer. Thus such ``early evaluation'' will
never be allowed in a secured circuit and computations will always be performed upon the
\textit{rendez-vous} of all data and \textbf{Acknowledge} inputs.

As the arrival of a new value is always signalled by a single wire changing value, the
parity of the Hamming weight of any signal changes each time a new value is transmitted.

Under a two-phase protocol, a gate will be ready to compute its output when all its inputs
show a parity opposed to the current output parity.

Under a \QPHP a gate is ready to compute as soon as each input has left then\INV state.  As
\INV is coded as $(0,0,\ldots,0)$ is has an even parity while any valid value, signalled
by a single wire at 1, has an even parity. The behaviour  of the gates under a \QPHP is
thus coherent with the one of the gates under \DPHP[s]. This will be useful for the design
of the FPGA.

\subsection{Acknowledge Signalling}
\label{discuss-OR-XOR}

The \ACK signal consists of a single wire, carrying a \{ \INV, \VAL\} under a \QPHP or an
\{ \Odd, \Even\} ``phase'' information, under a \DPHP.

Given the ``parity'' property of the signals, the \ACK signal is computed as the XOR of
all wires carrying the output signal. An OR gate would be enough under a
\QPHP. However it is easy to show the OR and the XOR functions are identical on
the allowed domain of values of the wires under a \QPHP.

This signal is sent by a given gate to those which drive its inputs.
When the output of a gate $S$ is sent to more than one gate, $D_1$, $D_2$,...,
a \RV is computed to combine the synchronization signals coming from
the $D_i$ into a single signal, fed to $S$.

\subsection{C-Element}

The \CE is the gate which implements the \RV of signals. It has an arbitrary
number $p$ of input wires, denoted $I_1$, $I_2$,\ldots,$I_p$, and a single output $Z$,
whose equations are:
\begin{equation} \label{c-element-1-eq}
Z = \begin{cases}
	1 &\EQIF I_1=I_2=\cdots=I_p=1 \\
	0 &\EQIF I_1=I_2=\cdots=I_p=0 \\
	Z &\EQOT.
	\end{cases}
\end{equation}
\noindent \REFEQ{c-element-2-eq} shows an equivalent form of \REFEQ{c-element-1-eq}.
\begin{equation}\label{c-element-2-eq}
Z = ( Z \wedge ( I_1 \vee \cdots \vee I_p ) ) \vee ( I_1 \wedge \cdots \wedge I_p ) \,.
\end{equation}
\noindent
Where $\wedge$ and $\vee$ are respectively the AND and OR operators.


 \begin{figure}
  \begin{center}
   \input{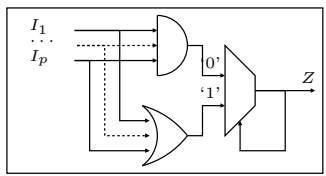} 
   \caption{{C-element implemented with a MUX}.}
   \label{07-c-element}
  \end{center}
 \end{figure}

\REFFIG{07-c-element} depicts the implementation of a C-element derived form
\REFEQ{c-element-2-eq}, using a multiplexer (MUX), which we use in out FPGA.

In an FPGA the C-element can be implemented in many ways.  A $p$-input C-element can be
implemented in $p+1$-input  \LUT, provided the output of the  \LUT can be fed back to one of
the inputs.


If $Z=0$, the MUX selects the AND gate, which will output 1 if and only if $\forall i
\in [1, p], I_i=1$.  When this condition becomes true, the output of the MUX becomes 1
and the output of the OR is selected to be sent to $Z$ instead of the one of the
AND.  As $\forall i, I_i=1 \Rightarrow \exists i : I_i\not=0$, the output is stable at
1. The output remains 1 until all inputs are back to 0.  \textit{Mutatis mutandis}
the same proof shows that the output of the gate comes back to 0 when all inputs are 0
and that this value is stable until all inputs are 1 again.  Thus the gate correctly
implements the \textit{rendez-vous} with no glitch.

\subsection{Asynchronous Computation \& Security}

\subsubsection{Timing Attack}

As each gate always waits for every input to be ready before computing its result, the duration
of the computations is independent of the data. However a dependency can be generated if
the lengths of the wires $x_i$ which implement a signal $x$ are different, thus generating
different propagation times for each value of $x$.

Thus the following necessary condition must hold: \textbf{for any pair
of gates} $(S,R)$, connected by a signal $x$, composed of wires $(x_0,x_1,\ldots,x_p)$:
\begin{itemize}
\item under the \QPHP, the propagation time of the transition from \INV to any
value and of the transition from any valid value to \INV $S$ to $R$ must be independent of the value;
\item under a \DPHP, the rising and falling times of any output wire must be
equal and independent from the former and next value of the signal.
\end{itemize}

As the condition must be fulfilled by any signal routed through the FPGA, this implies
that:
\begin{itemize}
\item in any routing channel, all wires must have the same length and the same capacity
with respect to \VDD or \VSS,
\item for any pair of wires in two routing channels connected by a switchbox, the
propagation time through the switchbox must be the same for all possible pairs,
\item for any input of a PLB, the propagation time from the network to the processing
elements must be uniform,
\item for any output, the propagation time to the routing network must be uniform.
\end{itemize}

If all these conditions are satisfied and if all PLB process information at the same speed
the \textit{timing attack}
\cite{kocher96timing,ches2000:timing-RSA-CRT,dhem98practical,hachez99timing,schindler-unleashing}
is impossible.

\subsubsection{Measurement of Power Consumption}

Under the \QPHP, two valid values are separated in time by a \INV value, implemented as
all wires at 0. The transition from \INV to a valid value $i$ consist in a rising edge
$0\mapsto1$ of wire $i$ and the return to \INV is the opposite falling transition.

In order to thwart these attacks the power consumption must be the same for the rising edge
of any of the wire $x_i$ which compose a signal $x$ and also for their falling edges. This
condition implies that lengths of the $x_i$ through the routing network be the same.

The necessary conditions to thwart the \textit{timing attack} are also necessary here but,
in addition the resistances of the output transistors must be equal.

\section{4-Phase Protocol}
\label{4-Phase-protocol}

This protocol is the simplest of all three because the instantaneous values of the wires
composing any signal are sufficient to determine the value of this signal. We will
implement the gates with:
\begin{itemize}
\item from 1 to 6 inputs, including the \ACKIN signals, and
\item from 2 to 4 outputs, \textbf{not} including the \ACKOUT signals.
\end {itemize}

\subsection{Encoding of Signals}

 \begin{figure}
  \begin{center}
   \input{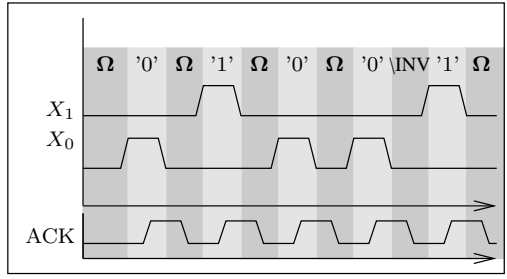} 
   \caption{{4-Phase Protocol}.}
   \label{dual-rail-states}
  \end{center}
 \end{figure}

Though it is not the only possible one, we shall use the one of \REFEQ{eq-4ph-encoding}
for a signal $x$ in the rest of this article:
\begin{equation}\label{eq-4ph-encoding}
	\begin{array}{@{\text{if }}l@{\,=\,}l@{:\ }l}
		 (x_0,x_1) & (0,0) &  x = \INV,                         \\
		 (x_0,x_1) & (1,0) &  x = \LZ,                          \\
		 (x_0,x_1) & (0,1) &  x = \LU\text{ and}                \\
		 ((x_0,x_1) & (1,1)& \text{\ \textbf{forbidden state}.} \\
	\end{array}
\end{equation}

\noindent The occurrence of the ``$(1,1)$''  \textbf{forbidden state} will always signal
either a malfunction or an attack against the system.
\REFFIG{dual-rail-states} depicts the succession of values on a signal $X$, represented by
2 wires $(x_1,x_0)$, and, when present, the associated transmissions of the ACK signal by
the receiver back to the transmitter.

\subsection{\OOF 2,2-input Gates}

Let $f(x,y):\gf\times\gf\mapsto\gf$ a two-variable Boolean function. Its output is a \OOF
2 signal represented by two wires $O_1$ and $O_0$. We denote respectively $f^1(x,y)$ and
$f^0(x,y)$ the functions computing the values of each wire.

\begin{figure}
  \begin{center}
   \input{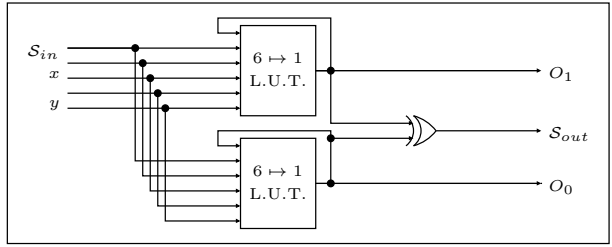} 
   \caption{{Minimal PLB for \QPH, 2-input gates}.}
   \label{2008-4ph-bin-2in}
  \end{center}
 \end{figure}

\REFFIG{2008-4ph-bin-2in} depicts the minimal structure of a PLB necessary to implement in
the most general way a gate with 2 binary inputs. Three signals enter the gate: 2 data
signals $x$ and $y$, respectively implemented by the $(x_0,x_1)$ and $(y_0,y_1)$ pairs of
wires, and \ACKIN, the synchronization signal.

The output value ($O$) is implemented by two \LUTS, respectively computing the $O_0$ and
$O_1$ wires. \REFEQ{eq-4ph-2in-simple} shows the equations of the outputs. In this
equation,
\begin{equation}
 \begin{array}{l@{\,=\,}l}
  O_1 &
   \begin{cases}
    f^1(x,y)&\EQIF(x\not=\INV)\wedge(y\not=\INV)\wedge(\ACKIN=0)\\
    0       &\EQIF(x    =\INV)\wedge(y    =\INV)\wedge(\ACKIN=1)\\
    O_1     &\EQOT.\\
   \end{cases}
  \\
  O_0 &
   \begin{cases}
    f^0(x,y)&\EQIF(x\not=\INV)\wedge(y\not=\INV)\wedge(\ACKIN=0)\\
    0       &\EQIF(x    =\INV)\wedge(y    =\INV)\wedge(\ACKIN=1)\\
    O_1     &\EQOT.\\
   \end{cases}
  \\
  \ACKOUT & O_0 \oplus O_1.
 \end{array}
 \label{eq-4ph-2in-simple}
\end{equation}

The ``memory effect'' implied by \REFEQ{eq-4ph-2in-simple} is implemented by sending each of
$O_0$ and $O_1$ to an input of the \LUT which drives it.  Thus the minimal practical size
for the \LUT is 64 bits, which can implement any 6-bit $\mapsto$ 1-bit function. As there
are two output bits the minimal size of the PLB is 2  \LUT.

Even if an OR gate would be enough, the \ACKOUT signal is computed by a XOR gate (See
\ref{discuss-OR-XOR}).

As the inputs to the  \LUT are the same, with the exception of the feedback wires, there
can be a single connection box to the routing network, which will divide by 2 the total
size of the connection boxes. \REFFIG{2008-4ph-bin-2in} shows the minimal structure of the
PLB, which allows to implement 2-input gates with synchronization.

\begin{remark}
In \REFEQ{2008-4ph-bin-2in}, each wire of $x$ and $y$ is loaded with exactly the same number
of inputs, as it is necessary to achieve the indiscernability of signals for a malevolent
observer.
\end{remark}

\subsection{\OOF 2, 3-input Gates}

\REFEQ{eq-4ph-2in-simple} can be immediately modified into \REFEQ{eq-4ph-3in} to add a third
input term $z$ and the new equation shows that we need a 7-input  \LUT with one feedback.

\begin{equation}
 \begin{array}{l@{\,=\,}l}
 O_1&
  \begin{cases}
   f^1(x,y,z)&\EQIF(x\not=\INV)\wedge(y\not=\INV)\wedge(z\not=\INV)\wedge(\ACKIN=0)\\
   0         &\EQIF(x    =\INV)\wedge(y    =\INV)\wedge(z    =\INV)\wedge(\ACKIN=1)\\
   O_1       &\EQOT.\\
  \end{cases}
 \\
 O_0 &
  \begin{cases}
   f^0(x,y,z)&\EQIF(x\not=\INV)\wedge(y\not=\INV)\wedge(z\not=\INV)\wedge(\ACKIN=0)\\
   0         &\EQIF(x    =\INV)\wedge(y    =\INV)\wedge(z    =\INV)\wedge(\ACKIN=1)\\
   O_0       &\EQOT.\\
  \end{cases}
 \\
 \ACKOUT & O_0 \oplus O_1.
 \end{array}
 \label{eq-4ph-3in}
\end{equation}

As the 3-input gates need 6 inputs for a 3-variable function, they cannot be implemented
in the structure of \REFFIG{2008-4ph-bin-2in}, on which each \LUTS has 5 inputs from the
routing network and 1 feedback input.

As it is not realistic to use two \LUTN{7}{1} because of the number of programming points
($2\times128$ bits), we separate the \textit{\RV + computation} function from the
\textit{memory} function and introduce a specific component: the \textbf{memory point}.

 \begin{figure}
  \begin{center}
   \input{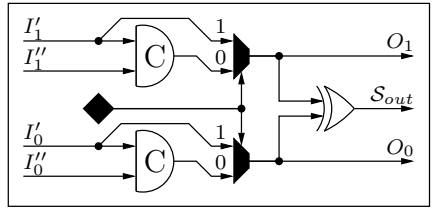} 
   \caption{{Memory Point}.}
   \label{2008-asyn-mem}
  \end{center}
 \end{figure}

\REFFIG{2008-asyn-mem} depicts the \textbf{memory point}, which consists in a pair of
\CE[s], together with a XOR gate, which computes the \ACKOUT signal.  Two MUX, under
control of a single programming point, allow to bypass the \CE[s]. It will be useful when
implementing the \DPHP[s].

 \begin{figure}
  \begin{center}
   \input{2008-4ph-bin-3in-b.pstex_t} 
   \caption{{Binary 3-input gate with \QPHP}.}
   \label{2008-4ph-bin-3in-b}
  \end{center}
 \end{figure}

\REFFIG{2008-4ph-bin-3in-b} depicts the schematic of the 2-input \OOF 3 gate. The   
ancillary ``return to \INV'' function is implemented
by a specialized 6-input OR gate while the \LUTS are programmed to compute the
\RV and the functions $F^1(x,y,z]$ and $f^1(x,y,z)$. 

\begin{remark}
Note that it is much better use of the \LUT than the one implied by
\REFFIG{2008-4ph-bin-2in}, in which all bits corresponding to the feedback input set to 1
are filled with '1' to implement the inclusive OR of all 4 input bits.
\end{remark}

The 4-\LUT PLB can implement two independent 3-input, \OOF 2 functions. Ex: a full-adder.

\begin{remark}
The wiring depicted by \REFFIG{2008-4ph-bin-3in-b} can handle any gate the inputs of
which sum up to 6 wires (Ex: one \ACKIN + one \OOF 2 input + one \OOF 3 input;
two \ACKIN, two \OOF 2 inputs, etc...).
\end{remark}

\begin{remark}
The feed-back and the associated MUX at the inputs of \LUT could be removed. However they
will be useful later for the implementation of the \DPHLP.
\end{remark}

\subsection{\OOF 3, 2-input Gates}

 \begin{figure}
  \begin{center}
   \input{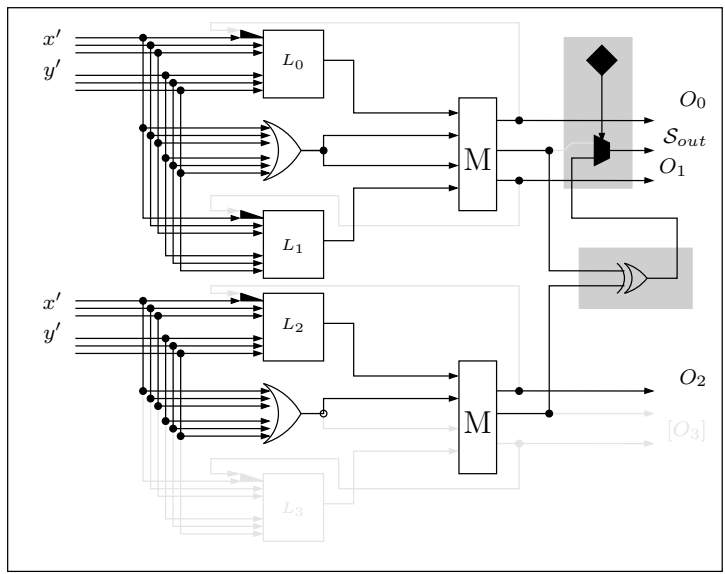} 
   \caption{{Structure of PLB needed to implement a ternary 2-input gate}.}
   \label{2008-4ph-ter-2in}
  \end{center}
 \end{figure}

Just as the \OOF 2, 3-Input Gates, the \OOF 3, 2-input gates need 6 inputs but they need
three outputs, each of them equipped with a memory point.  Strictly speaking, a \OOF 3 gate
needs three \LUT, each of them implementing one of the functions $O_i=f^i(x,y), i=0,1,2$.

However as most of the gates in a design will still be binary, the PLB features four \LUTS.
One of them will remain unused and filled with 0 when implementing a \OOF 3 gate. The
computation of the \ACKOUT signal needs some specialized hardware.
\REFFIG{2008-4ph-ter-2in} depicts the new PLB needed for a 2-input \OOF 3 gate, with the
supplementary devices gate in a grey rectangle:
\begin{itemize}
\item a MUX, controlled by a programming point, which allows to use the PLB either as two
separate 2-binary input, binary output gates or a single combined gate and
\item a single XOR gate which computes the XOR of all four outputs of the memory points.
\end{itemize}

For the same reason of compatibility with the binary gates, the inputs to the pairs of
\LUT are split into two groups. The load to each of the 12 input wires is exactly the same,
thus equalizing the power consumptions of all possible transitions on inputs.

\begin{remark}
The OR gates which compute the ``return to \INV''  signal are not grouped but will compute
output the same value as their input are the same.
\end{remark}

\subsection{Conclusion as for the 4-Phase Protocol}

In order to implement 2- and 3-inputs gates under the \QPHP, the PLB must at least consist
of four \LUTN 6 1, named $L_0$, $L_1$, $L_2$ and $L_3$.  One input of each \LUT can be
replaced with a feedback signal equal to the output pin.

The schematic depicted on \REFFIG{2008-4ph-ter-2in} is general: it can implement any
gate with:
\begin{itemize}
\item inputs consisting in any combination of 6 wires or less, including the \ACKIN
signals, and
\item outputs consisting of any combination of 4 wires, \textbf{not} counting the \ACKOUT
signals: 2 binary outputs, with separate \ACKOUT signals, 1 ternary output with a single
acknowledge-out signal or 1 quaternary output with an \ACKOUT signal.
\end{itemize}

\section{2-Phase Protocols}
\label{2-Phase-protocol}

\subsection{Phase of a Signal}

Under the \DPHP[s] valid values of a signal are not separated by ``\INV'' markers.
However, as the arrival of a new value (possibly identical to the preceding one) is
indicated by the toggling a exactly one wire, the parity of the Hamming weight of the
wires which represent a signal toggles at each new data.

In the following pages, \textit{the phase of the signal $X$}, denoted ``\PTY{X}'',
is by definition, the parity of the Hamming weight of the wires representing $X$.

\begin{remark}
For \ACK signals, which consist in a single wire, the phase is equal to the value of the
wire itself. The name of an \ACK signal $A$ will thus be used instead of \PTY A.
\end{remark}

At the beginning of the computation, all wires are set to a known value. \DPHP[s] require
that, after initialization and before any computation is started, the parities of all
signals be the same, say even. A simple way of ensuring this even parity is to initialize
all wires to 0.

As the phase of a signal toggles with every new valid value, a given gate is ready to
compute its output when the phases of all ``data'' signals at its inputs are the same,
different from the current phase of the output and the phase of the \ACKIN signal, if
present, the same as the output phase.

After the gate has performed its computation, the phase of its outputs become the common
one of the data inputs and thus the \ACKOUT signal toggles.

\subsection{2-Phase, LEDR Protocol}

This protocol is referred to as ``level-encoded dual-rail'', or LEDR~\cite{LINDER-HARDEN:SYNC-2-ASYNC}.

\subsubsection{Transmission of a Signal}

 \begin{figure}
  \begin{center}
   \input{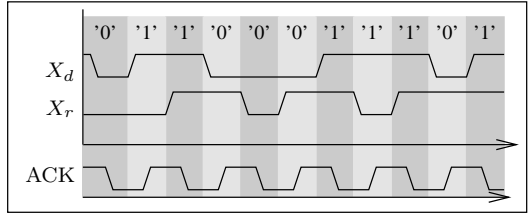} 
   \caption{{Transmission of a Signal and the acknowledge under the \DPHLP}.}
   \label{dual-rail-ledr}
  \end{center}
 \end{figure}

\REFFIG{dual-rail-ledr} shows the transmission protocol of the successive values of a
signal, together with the acknowledge signal. One can see that:
\begin{itemize}
\item
	a signal $X$ is represented by two wires: the  ``data wire'': ($X_d$) and
	 the ``repeat'' wire: ($X_r$);
\item
	each time a value is sent, exactly one wire toggles;
\item
	the value of the signal $X$ is the value of the $X_d$ signal, thus the oncoming a a
	new value, different from the preceding one is signalled by the toggling of $X_d$;
\item
	the oncoming of a new value, identical to the preceding one is signalled by $X_r$
	toggling; thus the instantaneous value of $X_r$ is irrelevant, only its toggling are
	significant.
\end{itemize}

\begin{remark}
The \DPHLP is restricted to binary signals. Otherwise, the transition between two values
would imply that more than a single wire toggle.
\end{remark}

\subsubsection{Binary 2-input Gates}

Let $f(x,y):\gf\times\gf\mapsto\gf$ a two-variable Boolean function.  The inputs are
represented by 4 wires: $x_d$, $x_r$, $y_d$ and $y_r$, to which a synchronization signal,
\ACKIN, may added and the output signal $O$ represented by two wires: $O_d$ and $O_r$,
together with an acknowledge output \ACKOUT.

The equations of the output wires are:
\begin{equation}
\begin{array}{l@{\,=\,}l}
O_d &
	\begin{cases}
		f{x}_d,y_d) & \EQIF (\PTY{x}) = 0) \wedge (\PTY{y})=0)\wedge(\ACKIN=1) ,\\
		f{x}_d,y_d) & \EQIF (\PTY{x}) = 1) \wedge (\PTY{y})=1)\wedge(\ACKIN=0) ,\\
		O_d        & \EQOT,\\
	\end{cases} \\
O_r &
	\begin{cases}
		 f{x}_d,y_d)           & \EQIF (\PTY{x}) = 0) \wedge (\PTY{y}) = 0 ) \wedge (\ACKIN=1),\\
		\overline{f{x}_d,y_d)} & \EQIF (\PTY{x}) = 1) \wedge (\PTY{y}) = 1 ) \wedge (\ACKIN=0),\\
		O_r                   & \EQOT.\\
	\end{cases} \\
\ACKOUT & O_d \oplus O_r.
\end{array}
\label{eq-2-ph-ledr-2-in}
\end{equation}

\REFEQ{eq-2-ph-ledr-2-in} shows that each of $(O_d,O_r)$ is a a function of 6 variables:
\begin{itemize}
\item two input data signals, represented by 4 wires,
\item one \ACKIN signal, represented by a single wire and
\item one feed-back signal, also 1 wire.
\end{itemize}

 \begin{figure}
  \begin{center}
   \input{2008-ledr-2-in.pstex_t} 
   \caption{{2-input Gate under the \DPHLP}.}
   \label{2008-ledr-2-in}
  \end{center}
 \end{figure}

These functions can be implemented in the same hardware as the corresponding gate under
the \QPHP. \REFFIG{2008-ledr-2-in} shows the assignment of the wires. 

The hardware elements which are not used to implement this gate are represented in dashed
lines:
\begin{itemize}
\item the $6^{th}$ input to the \LUTS, which is replaced by the feed-back,
\item the 6-input OR gate,
\item the memory element, which is programmed as ``transparent'' using its internal
programming point (See \REFFIG{2008-asyn-mem}).
\end{itemize}

Note that, opposite to the case of the \QPHP, here, the \ACKOUT value must be computed by
a XOR gate.

\subsubsection{3-input Gates}

Let $f(x,y,z):\gf^3\mapsto\gf$. \REFEQ{ledr-3in-eq} shows the expressions of the output
wires.
\begin{equation}\label{ledr-3in-eq}
\begin{array}{l@{\,=\,}l}
O_d&
 \begin{cases}
  f(x,y,z)
    & \EQIF(\PTY{x}=1)\wedge(\PTY{y}=1)\wedge(\PTY{z}=1)\wedge(\ACKIN=0),\\
  f(x,y,z)
    & \EQIF(\PTY{x}=0)\wedge(\PTY{y}=0)\wedge(\PTY{z}=0)\wedge(\ACKIN=1),\\
  O_d                      & \EQOT,
 \end{cases} \\
O_r &
 \begin{cases}
  f(x,y,z)
    & \EQIF(\PTY{x}=1)\wedge(\PTY{y}=1)\wedge(\PTY{z}=1)\wedge(\ACKIN=0),\\
  \overline{f(x,y,z)}
     & \EQIF(\PTY{x}=0)\wedge(\PTY{y}=0)\wedge(\PTY{z}=0)\wedge(\ACKIN=1),\\
  O_r
    & \EQOT.
 \end{cases} \\
\ACKOUT = O_d \oplus O_r \\
\end{array}
\end{equation}

\REFEQ{ledr-3in-eq} shows that each of $O_d$ and $O_r$ is a variable of 7 input variables
and cannot thus be implemented in a \LUTS.

\subsubsection{Practical Implementation}

 \begin{figure}
  \begin{center}
   \input{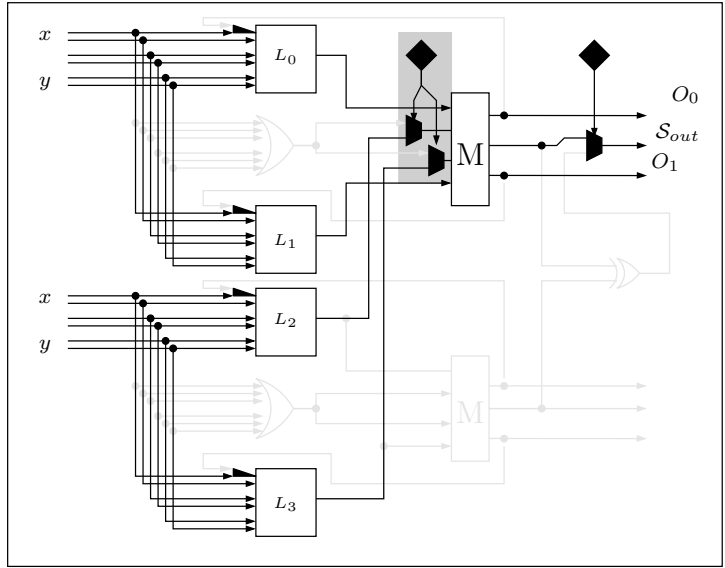} 
   \caption{{Implementation of the 3-input gate under the \DPHP}.}
   \label{2008-ledr-3-in}
  \end{center}
 \end{figure}

Under the \QPHP the outputs were set back to 0 by the \RV of the 0 coming from the \LUT
and the 0 coming from the 6-input OR gate. Under the \DPHP a OR gate cannot express the
``return to 0'' condition. Therefore the wiring of \REFFIG{2008-4ph-bin-3in-b} is modified
according to \REFFIG{2008-ledr-3-in}.

Two MUX, controlled by a programming point,  are added, which allow to replace the 6-in OR
gate by the two other \LUTS of the PLB. This way, each of $O_d$ and $O_r$ is now a \RV of
the outputs of 2 \LUT:

\begin{equation*}
\begin{array}{l@{\,=\,}l}
O_d & \RV( L_0 , L_2 ) \\
O_r & \RV( L_1 , L_3 ) \\
\end{array}
\end{equation*}

\REFEQ{ledr-2in-od-eq} shows the programming of \LUT $L_0$ and $L_2$ and
\REFEQ{ledr-2in-or-eq} shows the programming of \LUT $L_1$ and $L_3$.

\begin{equation}\label{ledr-2in-od-eq}
\begin{array}{l@{\,=\,}l}
L_0&
\begin{cases}
f(x_d,y_d,z_d) & \EQIF(\PTY{x}=0)\wedge(\PTY{y}=0)\wedge(\PTY{z}=0)\wedge(\ACKIN=1),\\
f(x_d,y_d,z_d) & \EQIF(\PTY{x}=1)\wedge(\PTY{y}=1)\wedge(\PTY{z}=1)\wedge(\ACKIN=0),\\
0 & \EQOT,
\end{cases} \\
L_2=&
\begin{cases}
f(x_d,y_d,z_d) & \EQIF(\PTY{x}=0)\wedge(\PTY{y}=0)\wedge(\PTY{z}=0)\wedge(\ACKIN=1),\\
f(x_d,y_d,z_d) & \EQIF(\PTY{x}=1)\wedge(\PTY{y}=1)\wedge(\PTY{z}=1)\wedge(\ACKIN=0),\\
1 & \EQOT.
\end{cases} \\
\end{array}
\end{equation}

When the conditions for a transition are fulfilled, $L_0$ and $L_2$ have the same value.
Thus the \RV occurs and $O_r$ takes its new value. Otherwise $L_0=0$ and $L_2=1$, the \CE
within the memory element has different values on its inputs and $O_d$ is locked.

\begin{equation}\label{ledr-2in-or-eq}
\begin{array}{l@{\,=\,}l}
L_1&
\begin{cases}
f(x_d,y_d,z_d) & \EQIF(\PTY{x}=0)\wedge(\PTY{y}=0)\wedge(\PTY{z}=0)\wedge(\ACKIN=1),\\
\overline{f(x_d,y_d,z_d)}
	& \EQIF(\PTY{x}=1)\wedge(\PTY{y}=1)\wedge(\PTY{z}=1)\wedge(\ACKIN=0),\\
0 & \EQOT,
\end{cases} \\
L_3=&
\begin{cases}
f(x_d,y_d,z_d) & \EQIF(\PTY{x}=0)\wedge(\PTY{y}=0)\wedge(\PTY{z}=0)\wedge(\ACKIN=1),\\
\overline{f(x_d,y_d,z_d)}
	& \EQIF(\PTY{x}=1)\wedge(\PTY{y}=1)\wedge(\PTY{z}=1)\wedge(\ACKIN=0),\\
1 & \EQOT.
\end{cases} \\
\end{array}
\end{equation}

\textit{Mutatis mutandis} the same demonstrations shows the validity of $O_r$.

\subsubsection{Conclusion on the 2-Phase, LEDR Protocol}

Apart from the shaded area in \REFFIG{2008-ledr-3-in} the \DPHLP needs the same resources
as the \QPHP.

As for security, all inputs to the gates have an equal load but  the value of a signal $X$
is the value of one of $x_d$. This is a potential security risk, which will have to be
investigated as soon as the ICs have been delivered.

\subsection{2-Phase, Edge Protocol}

 \begin{figure}
  \begin{center}
   \input{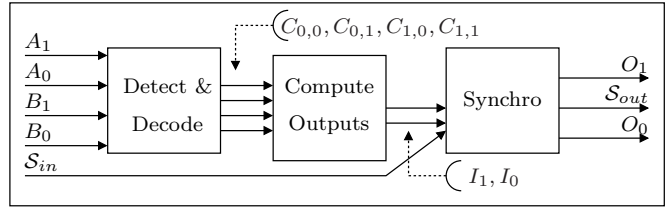} 
   \caption{{Global Structure of a 2-input 2-Phase-Edge gate}.}
   \label{2ph_e-2_in_gate}
  \end{center}
 \end{figure}

Signals under the \DPHEP can take an arbitrary number of values. Binary signals are
represented by 2 wires, ternary signals are represented by 3 wires, etc... However the
complexity of the gates is quadratic in the number of wires per signal. Thus the use of
this protocol is in practice limited to binary signals.  The complexity of the gates is
also quadratic in the number of inputs. Again this limits in practice the number of inputs
to 2.  In the sequel signals are binary and a signal $X$ is thus represented by 2 wires:
$(x_0,x_1)$.

The coding of the signals relies exclusively on toggling of wires. the instantaneous
values of the wires is always irrelevant. This means that the current state of 4 wires has
to be stored. Thus even for a 2-input gate all four \LUT of the PLB will have to be used.

\subsubsection{Structure of a Gate}

The global structure of a 2-input gate under the \DPHEP is depicted by
\REFFIG{2ph_e-2_in_gate}. The operation of the gate is divided in three steps:
\begin{description}
\item[Detection:      ] waits for an edge on $A_i$ and one on $B_j$ and toggles the
                        corresponding $C_{i,j}$,
\item[Computation:    ] toggles  $I_{f(i,j)}$ and
\item[Synchronization:] toggles $O_{f(i,j)}$ and \ACKOUT if and only if \ACKIN has toggled
                        since the last data output.
\end{description}

\paragraph{Detection: 2x2-decision wait}


 \begin{figure}
  \begin{center}
   \input{2phe-2x2.pstex_t} 
   \caption{{\TBTDW}.}
   \label{2phe-2x2}
  \end{center}
 \end{figure}

The detection and the decoding of the input data is performed by the circuitry known as
the ``\textbf{2x2-decision wait}'' or, shorter, the ``\TBTDW''. The circuitry, shown on
\REFFIG{2phe-2x2}, works as follows:
\begin{enumerate}
\item assume an initial state such that, for each \CE, the inputs are equal, (as
      this is the initial state, with all wires set to 0, the recurrence can start),
\item
	after an input value $i\in\{0,1\}$ has arrived on input port $A$ and an input value
	$j\in\{0,1\}$ on input port $B$, $A_i$ and $B_j$ have toggled (double-thickness
	continuous lines)
\item
	at this point:\\
	\begin{tabular}{@{- }l@{ to }ll@{$\ \Rightarrow\ $}ll}
	one input   & $C_{i,1-j}$&has toggled  & $C_{i,1-j}$& is unchanged,\\
	one input   & $C_{1-i,j}$&has toggled  & $C_{1-i,j}$& is unchanged,\\
	both inputs & $C_{i,j}$  &have toggled & $C_{i,j}$  & toggles,\\
	\end{tabular}
\item
	the new value of  $C_{i,j}$ is sent to the next stage and to the appropriate XOR gates
	to cancel the unwanted toggling of  $C_{i,1-j}$ and  $C_{i,1-j}$ (double-thickness
	dashed lines),
\item
	all four \CE[s] now have their inputs identical, which was the initial situation and
	$C_{i,j}$ has toggled, indicating to the next stage that:
		\begin{itemize}
		\item both input ports $A$ and $B$ have received a new data,
		\item the data just arrived on $A$ was $i$ and
		\item the data just arrived on $B$ was $j$.
	\end{itemize}
\end{enumerate}


Each of the $C_{i,j}$ can be expressed as:
\begin{equation}\label{c-ij-eq}
C_{i,j} = \RV( A_i \oplus C_{i,1-j} , B_j \oplus C_{1-i,j} )
\end{equation}

Each of the $C_{i,j}$ is a 5-term expression depending of:
\begin{itemize}
\item three feedback lines: $C_{i,j}$ (itself), $C_{i,1-j}$ and $C_{1-i,j}$ and
\item two input lines: $A_i$ and $B_j$.
\end{itemize}

Though the expression would fit in a \LUTS, the feedback from one \LUT to the other would
have to be routed through the general routing network, which has the following drawbacks:
\begin{itemize}
\item it consumes routing resources,
\item the timings of the feedbacks will be different between the feedback of a \LUT to
itself (which is routed inside the PLB) and other, routed outside. This could be an attack
point;
\item the 2-input gate will always need 2 PLB: one for the \TBTDW and one for
the computation itself.
\end{itemize}

 \begin{figure}
  \begin{center}
   \input{2008-all-feedbacks.pstex_t} 
   \caption{{PLB with all feedbacks for the \DPHEP}.}
   \label{2008-all-feedbacks}
  \end{center}
 \end{figure}

Therefore, these feedback have been added to the PLB, as shown on
\REFFIG{2008-all-feedbacks}, which. As on preceding figures, the black triangles at the
inputs of the \LUT are MUX controlled by programming points, which are denoted by
``\OU{.}{.}'' in \REFEQ{full-feedback-luts-eq}.

With these notations the equations of the 4 \LUTS are:

\begin{equation}\label{full-feedback-luts-eq}
\begin{array}{l@{\,\text{ =  \LUT(\,}\,}c@{\,,\,}c@{\,,\,}c@{\,,\,}c@{\,,\,}c@{\,,\,}c@{\,)}}
L_0&\OU{I' _0}{L_0}&\OU{I' _1}{L_1}&\OU{I' _2}{L_2}&\OU{I' _3}{L_3}&I' 4&I' 5)\\
L_1&\OU{I' _0}{L_0}&\OU{I' _1}{L_1}&\OU{I' _2}{L_2}&\OU{I' _3}{L_3}&I' 4&I' 5)\\
L_2&\OU{I''_0}{L_0}&\OU{I''_1}{L_1}&\OU{I''_2}{L_2}&\OU{I''_3}{L_3}&I''4&I''5)\\
L_3&\OU{I''_0}{L_0}&\OU{I''_1}{L_1}&\OU{I''_2}{L_2}&\OU{I''_3}{L_3}&I''4&I''5)\\
\end{array}
\end{equation}

 \begin{figure}
  \begin{center}
   \input{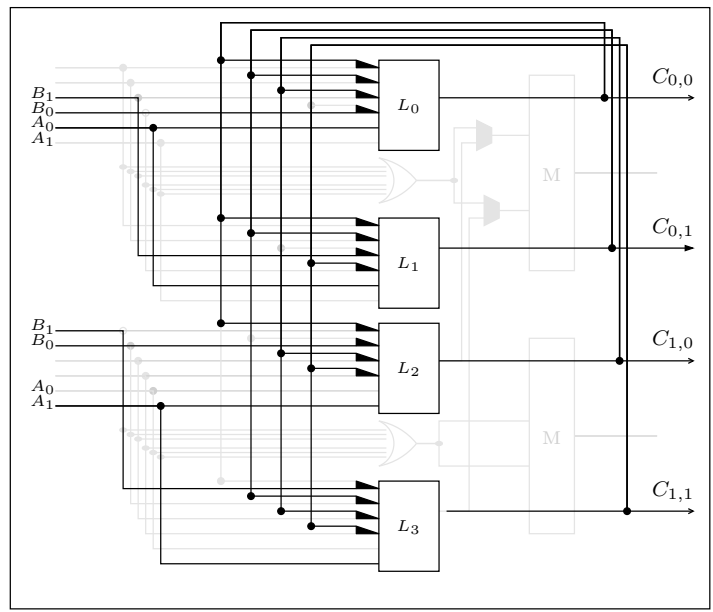} 
   \caption{{Wiring used to implement the $2\times1$-decision-wait}.}
   \label{2008-edge-all-2x2dw}
  \end{center}
 \end{figure}

To implement the \TBTDW the input lines are assigned as in \REFEQ{2x2-dw-pin-asgn-eq} and
depicted by \REFFIG{2008-edge-all-2x2dw}:
\begin{equation}\label{2x2-dw-pin-asgn-eq}
\begin{array}{
	l@{\,\rightarrow}l l@{\,\rightarrow}l l@{\,\rightarrow}l
	l@{\,\rightarrow}l l@{\,\rightarrow}l l@{\,\rightarrow}l
}
I' _0 & NC  & I' _1 & NC  & I' _2 & B_1 & I' _3 & B_0 & I' _4 & A_0 & I' _5 & A_1 \\
I''_0 & B_1 & I''_1 & B_0 & I''_2 & NC  & I''_3 & NC  & I''_4 & A_0 & I''_5 & A_1 \\
\end{array}
\end{equation}
in which ``$NC$'' means ``not connected'' and \REFEQ{2x2-dw-eq} shows the interconnection
of the feedbacks needed to implement the \TBTDW.

\begin{equation}\label{2x2-dw-eq}
\begin{array}{ l @{\,=\,}l @{\,=\,\LUT(\,}c @{\,,\,}c @{\,,\,}c @{\,,\,}c @{\,,\,}c @{\,,\,}c @{\,)} }
L_0 & C_{0,0} & C_{0,0} & C_{0,1} & C_{1,0} & B_0     & A_0   & A_1  \\
L_1 & C_{0,1} & C_{0,0} & C_{0,1} & B_1     & C_{1,1} & A_0   & A_1  \\
L_2 & C_{1,0} & C_{0,0} & B_0     & C_{1,0} & C_{1,1} & A_0   & A_1  \\
L_3 & C_{1,1} & B_1     & C_{0,1} & C_{1,0} & C_{1,1} & A_0   & A_1  \\
\end{array}
\end{equation}

\begin{remark}
$A_1$ is useless to compute $C_{0,0}$ and $C_{0,1}$ and $A_0$ is useless to compute
$C_{1,0}$ and $C_{1,1}$. The reason why these inputs are connected to the network but
ignored in the programming of the \LUT is that $B_0$ and $B_1$ are connected twice from
the network to the PLB and that the loads on this network must be identical for both
variables.
\end{remark}

\paragraph{Computation \& synchronization}

The \TBTDW stage provides a decoded output: $C_{i,j}$ toggles if $i$ and $j$ data have
arrived on inputs $A$ and $B$ respectively.

Computing the outputs is then straightforward: each of $O_1$ and $O_0$ outputs is the XOR
of the relevant $C_{i,j}$. Let's see some examples:

\begin{center}
\begin{tabular}{|l|l|l|}\hline
Gate & $O_1$ & $O_0$ \\\hline
AND  & $C_{1,1}                              $
     & $C_{0,0} \oplus C_{0,1} \oplus C_{1,0}$ \\\hline
NAND & $C_{0,0} \oplus C_{0,1} \oplus C_{1,0}$
     & $C_{1,1}                              $ \\\hline
OR   & $C_{1,1} \oplus C_{0,1} \oplus C_{1,0}$
     & $C_{0,0}                              $ \\\hline
NOR  & $C_{0,0}                              $
     & $C_{1,1} \oplus C_{0,1} \oplus C_{1,0}$ \\\hline
XOR  & $C_{0,1} \oplus C_{1,0}               $
     & $C_{0,0} \oplus C_{1,1}               $ \\\hline
NXOR & $C_{0,0} \oplus C_{1,1}               $
     & $C_{0,1} \oplus C_{1,0}               $ \\\hline
\end{tabular}
\end{center}

\paragraph{Synchronization}

 \begin{figure}
  \begin{center}
   \input{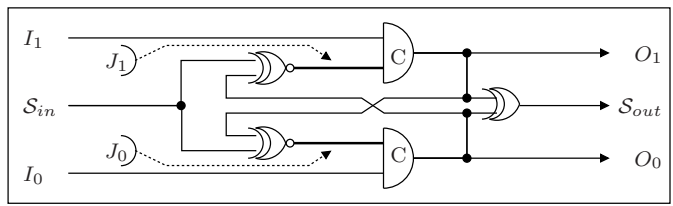} 
   \caption{{$2\times1$-decision-wait}.}
   \label{2phe-1x2}
  \end{center}
 \end{figure}


The synchronization is performed by a device called ``$2\times1$-decision-wait'' (or,
shorter: \TBODW).  \REFFIG{2phe-1x2} depicts the schematic of the \TBODW.

The \TBODW works as follows:
\begin{enumerate}
\item In the initial state, the following relations hold:
$O_1 = I_1$, $O_0=I_0$ and $\ACKIN=O_0\oplus O_1$, which imply $J_1 \not= I_1$ and
$J_0 \not= I_0$. (because $J_1 = \overline{\ACKIN \oplus O_0} = \overline{ O_0\oplus
O_1\oplus O_0} = \overline{O_1} = \overline{I_1}$, idem for $J_0$);
\item Assume $I_i$ toggles and thus becomes equal to $J_i$, the \CE transmits the common
value of its inputs to $O_i$,
\item as $O_i$ toggles, $J_{1-i}$ toggles too and becomes equal to $O_{1-j}$.
\item until \ACKIN toggles, we have $I_0 = J_0$ and $I_1 = J_1$: even if one of the
inputs toggles, the \CE[s] will remain stable;
\item when \ACKIN toggles, $J_0$ and $J_1$ toggle and the system is back in the initial
state.
\end{enumerate}


 \begin{figure}
  \begin{center}
   \input{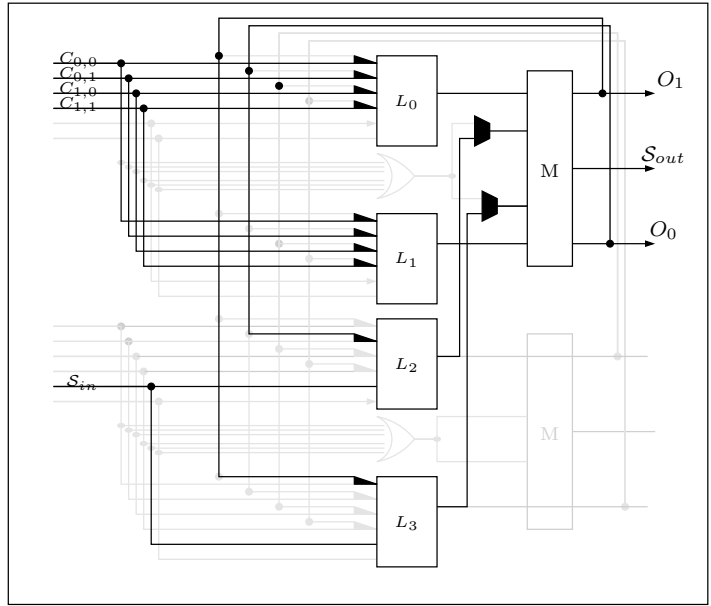} 
   \caption{{Wiring used to implement the $2\times1$-decision-wait}.}
   \label{2008-edge-all-2x1dw}
  \end{center}
 \end{figure}

If one wants to combine the computation stage with the \TBODW, it cannot be done in 2
\LUT.

It is not because of the complexity of the functions: each of $O_1$ and $O_0$ is a function
of 2 feed-backs, 1 \ACKIN and at most 3 $C_{i,j}$, at least if one does not want to
implement trivial, constant functions.

However, the set of 2 \LUT together would need 2 feed-backs, 1 \ACKIN and 4 $C_{i,j}$,
which is one more than the number of available wires. Thus we must use a full PLB.

If we use the full PLB, the memory element will provide the necessary \CE and the \LUT
become purely combinatorial. The inputs will be assigned following \REFEQ{edge-2x1-in-eq}
and depicted on  \REFFIG{2008-edge-all-2x1dw}:
\begin{equation}
\label{edge-2x1-in-eq}
(I'_0, I'_1,I'_2,I'_3)=( C_{0,0}, C_{0,1}, C_{1,0}, C_{1,1} ) \text{ and } I''_4 =\ACKIN
\end{equation}

Then the \LUT are programmed as by \REFEQ{edge-2x1-lut-eq}
\begin{equation}
\label{edge-2x1-lut-eq}
\begin{array}{l@{\ =\ }l}
L_0 & f^1( C_{0,0} , C_{0,1} , C_{1,0} , C_{1,1} ) \\
L_1 & f^0( C_{0,0} , C_{0,1} , C_{1,0} , C_{1,1} ) \\
L_2 & \overline{ O_0 \oplus \ACKIN } \\
L_3 & \overline{ O_1 \oplus \ACKIN } \\
\end{array}
\end{equation}

\subsubsection{Conclusion on the 2-Phase, Edge Protocol}

The \DPHEP is difficult to implement in a FPGA without special hardware added to the PLB:
it takes two PLB to implement a single 2-input gate.

However this protocol has advantages as for security because the instantaneous value of
the wires is not significant in itself. For instance \LU is represented alternatively by
the rising and the falling edge of a given wire. An attacker trying DPA, for instance,
would have to exhibit the difference between the average consumption of both edges on wire
'1' and the same average on wire '0'.

\section{Programming the FPGA}
\label{programming-the-fpga}

The FPGA can be partially programmed: it is divided in square blocks which can be
programmed separately from the other.

 \begin{figure}
  \begin{center}
   \input{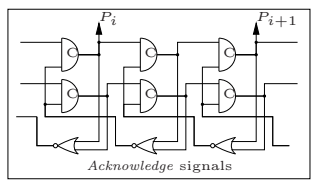} 
   \caption{{FIFO memory for programming}.}
   \label{2008-jmm-fifo-pg}
  \end{center}
 \end{figure}

The programming chain is a set of asynchronous FIFO memories. An elementary stage of these
FIFO is depicted by \REFFIG{2008-jmm-fifo-pg}.

At RESET time, all \CE[s] are set to zero by a general RESET wire. Then the programming
bits are fed to the FIFO, separated by \INV values. The last stage of each FIFO is
particular: the \ACK signal is controlled by an external pin. During the programming of
the block, the \ACK signal is held low. This way the programming bits are stacked in the
FIFO and the FPGA becomes functional.

If a partial reconfiguration is wanted, the chosen blocks are cleared by allowing the \ACK
signal of their last stage to acknowledge the value in the last stage. Then the FIFO is
activated again until all bits have gone thought it. At this point, the \ACK signal is
blocked again and the FIFO is ready to receive a new set of configuration bits.

During the configuration of the FPGA, all outputs of PLB are kept at 0 to avoid
short-circuits. The PLB are programmed first, while all switchboxes are left in an
insulation mode. Then the switchboxes are programmed to connect the newly reconfigured
part to be connected to the still working part. It is the designer's responsibility to
ensure that the new part can create no conflict with the existing part.

\section{Conclusion}
\label{conclusion}

We have presented the programmable logic block of an asynchronous FPGA, which is oriented towards security rather than
performance. In particular we have chosen not to implement one of the advantages of an
asynchronous design, which usually allows to compute in average time: the early evaluation.
This choice is deliberate as early evaluation is a security risk~\cite{suzuki-ches2006}.

The FPGA can accommodate various sizes of data as well as various styles of asynchronous control,
thus making it possible for the end user to design mixed styles of logic, depending on the applicative requirements.
Incidentally, this FPGA is also a valuable prototype that allows to perform comparisons between styles of asynchronous protocols.

A silicon is being manufactured and will be used for intensive testing. The different
resistances of the various protocols against SCA will be evaluated. In particular the
strict link under the \DPHLP between the value of a signal $X$ and the one of the $X_d$
wire will decide whether this protocol is suitable at all for a secure implementation.

\bibliographystyle{elsart-num}

\end{document}